\documentclass[10pt,twoside]{article}
\usepackage{amssymb}
\usepackage{amsmath}
\usepackage[dvips]{graphicx}
\usepackage[latin1]{inputenc}
\usepackage{pifont}

\begin{document}

\title{Study of the apsidal precession of the Physical Symmetrical Pendulum.}
\author{Héctor R. Maya$^{1,2}$\thanks{%
hrmaya@hotmail.com}, Rodolfo A. Diaz$^{2}$\thanks{%
radiazs@unal.edu.co}, William J. Herrera$^{2}$\thanks{%
jherreraw@unal.edu.co} \\
$^{1}$Universidad de Córdoba,\\
Departamento de Física. Montería, Colombia.\\
$^{2}$Universidad Nacional de Colombia,\\
Departamento de Física. Bogotá, Colombia.}
\maketitle

\begin{abstract}
We study the apsidal precession of a Physical Symmetrical Pendulum (Allais'
precession) as a generalization of the precession corresponding to the Ideal
Spherical Pendulum (Airy's Precession). Based on the Hamilton-Jacobi
formalism and using the technics of variation of parameters along with the
averaging method, we obtain approximate solutions, in terms of which the
motion of both systems admits a simple geometrical description. The method
developed in this paper is considerably simpler than the standard one in
terms of elliptical functions and the numerical agreement with the exact
solutions is excellent. In addition, the present procedure permits to show
clearly the origin of the Airy's and Allais' precession, as well as the
effect of the spin of the Physical Pendulum on the Allais' precession.
Further, the method can be extended to the study of the asymmetrical
pendulum in which an exact solution is not possible anymore.

\bigskip

\textbf{PACS}: 45.40.-f, 83.10.Ff, 45.20.Jj, 47.10.Df

\textbf{Keywords}: Ideal spherical pendulum, Physical symmetrical pendulum, Hamilton-Jacobi formalism, Averaging method, Apsidal precession.

\end{abstract}

\section{Introduction}

A physical symmetrical pendulum is a particular case of the symmetrical top
in which the center of mass (CM) is located below the fixed point, and the
precession and spin can be considered as small perturbations with respect to
the nutation\footnote{%
In the most usual scenario of the symmetrical top, the spin is the dominant
part of its motion.}. Lagrange \cite{Lagrange}; Poisson \cite{Poisson};
Golubev \cite{Golubev} and Leimanis \cite{Leimanis}, found exact solutions
(in terms of elliptical functions) for the dynamics and kinematics of the
symmetrical top under the action of gravity. On the other hand, Johansen and
Kane \cite{Johansen} obtained an approximate solution for the Ideal
Spherical Pendulum (ISP) using the method of averaging by using canonical
variables. In addition, Miles \cite{Miles} analyzed the response of the ISP
under a harmonic excitation, while Hemp and Sethana \cite{Hemp} studied the
dynamics of the ISP when the support undergoes vertical motion.

More closely related with this paper are the articles of Airy \cite{Airy},
Olsson \cite{Olsson} and Synge \cite{Synge}. By using different methods of
approximation, these authors found the angular frecuency of precession that
undergoes the apsidal axis of the projection of the ISP trajectory (the
so-called Airy's precession). In these studies it is assumed that the motion
starts with small initial amplitudes. More recently, Gusev, Rudenko and
Vinogradov \cite{Gusev}, considered this precession when the pendulum is
submitted to small perturbations coming from the anisotropy of the support
and they found an analytical formula for the angular frequency of the plane
of oscillation as a function of the initial conditions and the anisotropy of
the support.

As for the Physical Symmetrical Pendulum (PSP), the most usual analysis of
its dynamics is greatly simplified due to the assumption of spin dominance,
in which the nutation and precession are considered small perturbations \cite%
{Goldstein}. By contrast, we study the symmetrical top (i.e. the PSP)\
considering it as a pendulum with a fixed point, in a regime of nutation
dominance. We shall assume that the pendulum is released near to the surface
of the earth with small initial amplitudes and a small transverse initial
velocity, but without initial spin (with respect to the earth). On the other
hand, in an inertial reference frame the PSP has a small correction to the
initial precession and spin (owing to the rotation of the earth), but they
are very small (of the order of $10^{-4}$\ rad/seg) and are kept small at
all times, from which the nutation becomes the dominant motion. In this
regime of initial conditions the PSP describes trajectories that are
approximately elliptical and that precess very slowly. In the case of the
ISP this effect is called \textquotedblleft Airy's
precession\textquotedblright\ while in the case of a physical pendulum
(which is our case) we shall call it \textquotedblleft Allais'
precession\textquotedblright\ \cite{Allais}. The results of this study
represent a first approximation to the dynamics of the paraconical pendulum,
originally designed by Allais, and currently used widely by many researchers
in the characterization of gravitational anomalies during eclipses \cite%
{Goodey}. As we shall see, the spin introduces a significant correction to
the precession of the PSP (with respect to the precession of the ISP).

Our paper is distributed as follows: In section \ref{sec:ISP}, we study the
Ideal Spherical Pendulum, using the Hamilton-Jacobi approach combined with
the technics of variation of parameters and the averaging method. These
approximate results are numerically compared with the exact solution showing
an excellent agreement between them even for long times. In this approach,
the Airy's precession appears naturally. The methods developed in this
section are applied to the symmetrical physical pendulum in section \ref%
{sec:PSP}, and once again an excellent agreement with the exact solution
even for large times is apparent. The corresponding Allais' precession
appears clearly from the formalism, as well as the correction to this
precession coming from the spin. Section \ref{sec:conclusions} shows our
conclusions and appendix \ref{sec:appendix} shows some few technical details.

\section{Apsidal precession in the ideal spherical pendulum\label{sec:ISP}}

We now study the origin of the Apsidal precession of the ideal spherical
pendulum, keeping in mind that our aim is just to establish the general
framework to apply a similar procedure for the physical symmetrical
pendulum. We assume that the (small) initial amplitudes are of the order of $%
0.1~$rad (experimental conditions).

\begin{figure}[tbh]
\begin{center}
\includegraphics[scale=0.7]{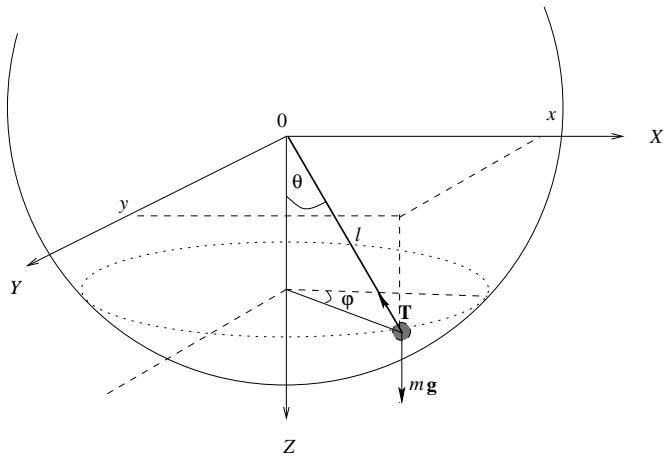}
\end{center}
\caption{\emph{Inertial system }$\emph{XYZ}$\emph{\ and the spherical
pendulum. The angular coordinates $\protect\theta $ and$\ \protect\varphi \
are\ shown,\ as\ well\ as\ the\ weight\ and\ tension$.}}
\label{Fig1}
\end{figure}

Let us consider a system of axes $XYZ$ fixed on an inertial reference frame
and the spherical pendulum of mass $m$ and length $l,$ as displayed in Fig. %
\ref{Fig1}

The natural coordinates that exhibit the symmetries of the IPS are the
spherical coordinates $\theta $ and $\varphi $. The Lagrangian in such
coordinates becomes%
\begin{equation}
L=\frac{1}{2}ml^{2}\left( \dot{\theta}^{2}+\dot{\varphi}^{2}\sin ^{2}\theta
\right) +mgl\cos \theta .  \label{1.153}
\end{equation}%
The associated canonically conjugate momenta are given by%
\begin{equation}
p_{\theta }=\frac{\partial L}{\partial \dot{\theta}}=ml^{2}\dot{\theta}\ ,\
\ p_{\varphi }=ml^{2}\dot{\varphi}\sin ^{2}\theta ,  \label{1.154}
\end{equation}%
since $\varphi $ is cyclic, its conjugate momentum $p_{\varphi }$ is
constant and can be identified as the $z-$component of the angular momentum.
The Hamiltonian of the system reads 
\begin{subequations}
\label{1.155}
\begin{gather}
h=\dot{\theta}p_{\theta }+\dot{\varphi}\ p_{\varphi }-L,  \label{1.155a} \\
h=\frac{1}{2ml^{2}}\left( p_{\theta }^{2}+\frac{p_{\varphi }^{2}}{\sin
^{2}\theta }\right) -mgl\cos \theta ,  \label{1.55b}
\end{gather}%
or in dimensionless units 
\end{subequations}
\begin{equation}
H=\frac{1}{2}\left( P_{\theta }^{2}+\frac{P_{\varphi }^{2}}{\sin ^{2}\theta }%
\right) -\cos \theta ,  \label{1.156}
\end{equation}%
where we have introduced the following definitions%
\begin{equation}
H\equiv \frac{h}{mgl}\ ,\ \ \omega _{0}\equiv \sqrt{\frac{g}{l}}\ ,\ \ \tau
\equiv \omega _{0}t\ ,\ \ P_{\theta }\equiv \frac{p_{\theta }}{ml^{2}\omega
_{0}}\ ,\ \ P_{\varphi }\equiv \frac{p_{\varphi }}{ml^{2}\omega _{0}}.
\label{1.157}
\end{equation}

Now we introduce a canonical transformation: $\left( \theta ,\varphi
,P_{\theta },P_{\varphi }\right) \rightarrow \left( \bar{\theta},\varphi ,%
\bar{P}_{\theta },P_{\varphi }\right) $ which keeps unaltered the variables
associated with the precession (so that the constant of motion is still
apparent), and permits to eliminate the circular functions from this
Hamiltonian. An appropriate generating function of type II\ \cite{Goldstein}
for this Canonical Transformation (CT)\ reads%
\begin{equation}
F_{2}(\theta ,\varphi ,\bar{P}_{\theta },P_{\varphi })=\bar{P}_{\theta }\sin
\theta +\bar{P}_{\varphi }\varphi .  \label{tcp}
\end{equation}%
The formulas of transformation\ for the variables that describe the nutation
are given by 
\begin{equation}
\bar{\theta}=\frac{\partial F_{2}}{\partial \bar{P}_{\theta }}=\sin \theta \
,\ \ P_{\theta }=\frac{\partial F_{2}}{\partial \theta }=\bar{P}_{\theta
}\cos \theta ,  \label{tc1}
\end{equation}%
while the new variables that describe the precession are identical to the
old ones. Therefore, we continue using the same symbols $\varphi $ and $%
P_{\varphi }$ for them. Using this result in Eq. (\ref{1.156}) we obtain the
new Hamiltonian%
\begin{equation}
\bar{H}=\frac{1}{2}\left[ \bar{P}_{\theta }^{2}(1-\bar{\theta}^{2})+\frac{%
P_{\varphi }^{2}}{\bar{\theta}^{2}}\right] -\sqrt{1-\bar{\theta}^{2}}.
\label{hb}
\end{equation}

Now, since we are interested in a regime of small initial amplitudes (i.e.
small values of $\theta _{0}$) and since $\bar{\theta}\equiv \sin \theta $,
we also have small values of the coordinate $\bar{\theta}$. Consequently, we
can expand the radical in power series and keep terms up to fourth order in $%
\bar{\theta}$. We do this because this is the lowest order in which the
apsidal precession appears. By doing such an expansion and omitting constant
terms in the Hamiltonian we obtain%
\begin{equation}
\bar{H}=\frac{\bar{P}_{\theta }^{2}}{2}+\frac{\bar{\theta}^{2}}{2}+\frac{%
P_{\varphi }^{2}}{2\bar{\theta}^{2}}-\frac{\bar{P}_{\theta }^{2}\bar{\theta}%
^{2}}{2}+\frac{\bar{\theta}^{4}}{8},  \label{1.158}
\end{equation}%
We identify the terms up to second order in the momenta and/or variables
with the non-perturbed Hamiltonian (let us recall that all momenta and
variables are dimensionless), while higher order terms are identified with
the Hamiltonian of perturbation 
\begin{subequations}
\label{1.159}
\begin{gather}
H_{0}=\frac{1}{2}\bar{P}_{\theta }^{2}+\frac{\bar{\theta}^{2}}{2}+\frac{%
P_{\varphi }^{2}}{2\bar{\theta}^{2}},  \label{1.159a} \\
H_{1}=-\frac{\bar{P}_{\theta }^{2}\bar{\theta}^{2}}{2}+\frac{\bar{\theta}^{4}%
}{8}.  \label{1.159b}
\end{gather}

We shall use these Hamiltonians to determine the approximate solutions for
the ISP. To do it, we shall apply the Hamilton-Jacobi formalism along with
the technics of variation of parameters as well as the method of averaging.

\subsection{Restricted Hamilton-Jacobi Method for the non-perturbed
Hamiltonian}

We start by finding the exact solution for the Hamiltonian $H_{0}$ by means
of the implementation of the restricted Hamilton-Jacobi (RHJ) method, that
is for the Hamilton's Characteristic Function \cite{Goldstein}. Equation (%
\ref{1.159a}) shows that $\varphi $ is cyclic in $H_{0}$, hence we can write
the Hamilton's Characterisitc Function $W$ in the form 
\end{subequations}
\begin{equation}
W=W_{\bar{\theta}}+\alpha _{2}\varphi ,  \label{PEI-2}
\end{equation}%
where $P_{\varphi }\equiv \alpha _{2}$ is the canonical conjugate momentum
associated with $\varphi ,$\ and the RHJ equation for $H_{0}$ becomes%
\begin{equation}
\frac{1}{2}\left( \frac{\partial W}{\partial \bar{\theta}}\right) ^{2}+\frac{%
\bar{\theta}^{2}}{2}+\frac{\alpha _{2}^{2}}{2\bar{\theta}^{2}}=\alpha _{1},
\label{pei-3}
\end{equation}%
where $\alpha _{1}$ corresponds to the numerical value of $H_{0}$ i.e. the
mechanical energy of the system described by this Hamiltonian 
\begin{equation}
\alpha _{1}=\frac{1}{2}\bar{P}_{\theta 0}^{2}+\frac{\bar{\theta}_{0}^{2}}{2}+%
\frac{\alpha _{2}^{2}}{2\bar{\theta}_{0}^{2}}  \label{alfa1}
\end{equation}%
Equation (\ref{pei-3}) is an ordinary differential equation for $W_{\bar{%
\theta}}$ that can be solved directly 
\begin{equation}
\frac{dW_{\bar{\theta}}}{d\bar{\theta}}=\pm \frac{\sqrt{-\alpha
_{2}^{2}+2\alpha _{1}\bar{\theta}^{2}-\bar{\theta}^{4}}}{\bar{\theta}},
\label{PEI-4}
\end{equation}%
and the function $W$ given by (\ref{PEI-2}) becomes%
\begin{equation}
W=\alpha _{2}\varphi \pm \int \frac{\sqrt{-\alpha _{2}^{2}+2\alpha _{1}\bar{%
\theta}^{2}-\bar{\theta}^{4}}}{\bar{\theta}}d\bar{\theta}.  \label{PEI-5}
\end{equation}%
The new angular variables are obtained from%
\begin{eqnarray}
\tau +\beta _{1} &=&\frac{\partial W}{\partial \alpha _{1}}=\pm \int \frac{%
\bar{\theta}}{\sqrt{-\alpha _{2}^{2}+2\alpha _{1}\bar{\theta}^{2}-\bar{\theta%
}^{4}}}d\bar{\theta}=\mp \frac{1}{2}\arcsin \left( \frac{\alpha _{1}-\bar{%
\theta}^{2}}{k}\right) ,  \label{PEI-6} \\
k^{2} &\equiv &\alpha _{1}^{2}-\alpha _{2}^{2}  \label{PEI-7}
\end{eqnarray}%
the initial conditions that we shall consider [Equations (\ref{1.163b}) with 
$\theta _{0}<<1~$Rad], combined with Eq. (\ref{alfa1}) lead to $\alpha
_{1}^{2}>\alpha _{2}^{2}$, and $0<\theta <\pi /2$. The first condition says
that $k^{2}$ is a positive constant while the second condition leads us to
preserve only the upper sign in Eqs. (\ref{PEI-6}). Further, the sign of the
integral associated with the equation for $\partial W/\partial \alpha _{2}$
is chosen accordingly%
\begin{equation}
\beta _{2}=\frac{\partial W}{\partial \alpha _{2}}=\varphi -\int \frac{%
\alpha _{2}}{\bar{\theta}\sqrt{-\alpha _{2}+2\alpha _{1}\bar{\theta}^{2}-%
\bar{\theta}^{4}}}d\bar{\theta},  \label{pei-8}
\end{equation}%
in order to evaluate this integral we use Eq. (\ref{PEI-6}) to define the
variable%
\begin{equation}
u=\tau +\beta _{1}=-\frac{1}{2}\arcsin \left( \frac{\alpha _{1}-\bar{\theta}%
^{2}}{k}\right)   \label{u+tbeta1}
\end{equation}%
substituting (\ref{u+tbeta1}) in (\ref{pei-8}) we can write $\varphi $ as 
\begin{subequations}
\label{1.160}
\begin{gather}
\varphi =\beta _{2}+\int \frac{\alpha _{2}}{\alpha _{1}+k\sin (2u)}du=\beta
_{2}+\arctan \left[ \frac{1}{\alpha _{2}}\left( k+\alpha _{1}\tan u\right) %
\right]   \label{1.160a} \\
\varphi =\beta _{2}+\arctan \left\{ \frac{1}{\alpha _{2}}\left[ k+\alpha
_{1}\tan \left( \tau +\beta _{1}\right) \right] \right\}   \label{1.160b}
\end{gather}

From Eqs. (\ref{PEI-6}), (\ref{PEI-4}) and (\ref{1.160}) we have 
\end{subequations}
\begin{subequations}
\label{PEI-9}
\begin{gather}
\bar{\theta}=\sqrt{\alpha _{1}+k\sin \left[ 2\left( \beta _{1}+\tau \right) %
\right] },  \label{PEI-9a} \\
\bar{P}_{\theta }=\frac{dW_{\bar{\theta}}}{d\bar{\theta}}=\frac{k\cos \left[
2\left( \beta _{1}+\tau \right) \right] }{\bar{\theta}},  \label{PEI-9b} \\
\varphi =\beta _{2}+\arctan \left\{ \frac{1}{\alpha _{2}}\left[ k+\alpha
_{1}\tan \left( \beta _{1}+\tau \right) \right] \right\} ,  \label{PEI-9c}
\end{gather}%
these equations along with $P_{\varphi }=\alpha _{2},$ form the solution for
the non-perturbed Hamiltonian $H_{0}$.

\subsection{Method of averaging using canonical variables for the complete
Hamiltonian}

The next step is to obtain an approximate solution for the complete
Hamiltonian $\bar{H}\ $of Eq. (\ref{1.158}), based on the exact solution (%
\ref{PEI-9}) for the non-perturbed Hamiltonian $H_{0}$. To do this, we shall
use the method of averaging using canonical variables. Our approach is a
variation of the approximation proposed by K. F. Johansen and T. R. Kane 
\cite{Johansen}. The goal of such an approach is to obtain a set of
approximate equations of easy solution for the canonical variables, by
applying the averaging method in the version proposed by
Krylov-Bogoliubov-Mitropolsky \cite{Krylov}. As we shall see, this technics
is based on the variation of parameters and \textquotedblleft the fast
integration\textquotedblright .

In the framework of the Hamilton-Jacobi equation for Hamilton's Principal
Function, it is clear that the transformations (\ref{PEI-9}) are associated
with a generating function of type II given by 
\end{subequations}
\begin{equation}
S_{0}=S_{0}(\bar{\theta},\varphi ,\alpha _{1},\alpha _{2},\tau )=W(\bar{%
\theta},\varphi ,\alpha _{1},\alpha _{2})-\alpha _{1}\tau 
\end{equation}%
such that%
\begin{equation}
P_{j}=\frac{\partial S_{0}}{\partial q_{j}}\ ,\ \ \beta _{j}=\frac{\partial
S_{0}}{\partial \alpha _{j}}.
\end{equation}%
Note that at this step we are using the Hamilton-Jacobi formalism for the
Hamilton's principal function $S$. This is a natural choice since at this
moment what we pretend is to see the way in which $H_{1}$ modifies the exact
solution (\ref{PEI-9}) of $H_{0}$. We can do this by demanding that the
associated CT reduces the total Hamiltonian $H_{0}+H_{1}$ to the Hamiltonian 
$H_{1}$. Of course, it means that the new Hamiltonian must be numerically
different from the old one, and it is possible only if the generating
function depends on time explicitly as is the case of $S_{0}.$ We can then
propose a generating function\ $S$ for a canonical transformation to a new
set of canonical variables $Q_{1},Q_{2},P_{1}$ and $P_{2}$ if we replace in $%
S_{0}$ the constants $\alpha _{1}\ $and $\alpha _{2}$ by $P_{1}$ and $P_{2}$
respectively, that is%
\begin{equation}
S=S_{0}\left( \bar{\theta},\varphi ,P_{1},P_{2},\tau \right) ,
\label{S0 this trasnform}
\end{equation}%
where now $P_{i}$ and $Q_{i}$ have become variables (this is the method of
variation of parameters). The transformations induced by (\ref{S0 this
trasnform}) are obviously (\ref{PEI-9}) but with the replacements $\alpha
_{i}\rightarrow P_{i},\ \beta _{i}\rightarrow Q_{i}$%
\begin{subequations}
\label{PEI-10}
\begin{gather}
\bar{\theta}=\sqrt{P_{1}+k\sin \left[ 2\left( Q_{1}+\tau \right) \right] },
\label{PEI-10a} \\
\bar{P}_{\theta }=\frac{k\cos \left[ 2\left( Q_{1}+\tau \right) \right] }{%
\sqrt{P_{1}+k\sin \left[ 2\left( Q_{1}+\tau \right) \right] }}\ ,\ \
P_{\varphi }=P_{2}  \label{PEI-10b} \\
\varphi =Q_{2}+\arctan \left[ \frac{1}{P_{2}}\left( k+P_{1}\tan \left(
Q_{1}+\tau \right) \right) \right] ,  \label{PEI-10c}
\end{gather}%
where 
\end{subequations}
\begin{equation}
\ k=\sqrt{P_{1}^{2}-P_{2}^{2}}.  \label{PEI-11}
\end{equation}%
The new Hamiltonian $K$ as a function of $Q_{1},\ Q_{2},\ P_{1}$ and $P_{2}$
is given by%
\begin{equation}
K=H_{0}+H_{1}+\frac{\partial S}{\partial t}=H_{1}\left( Q,P\right) ,
\label{PEI-12}
\end{equation}%
where we have used $H_{0}=P_{1}$ and $\partial S/\partial t=-P_{1}.$
Expressing $H_{1}$ [Eq.\ (\ref{1.159b})] in terms of these new variables,
yields 
\begin{subequations}
\label{PEI-13}
\begin{gather}
K=-\frac{\bar{P}_{\theta }^{2}\bar{\theta}^{2}}{2}+\frac{\bar{\theta}^{4}}{8}
\label{PEI-13a} \\
K=\frac{1}{8}\left\{ P_{1}+k\sin \left[ 2\left( Q_{1}+\tau \right) \right]
\right\} ^{2}-\frac{1}{2}k^{2}\cos ^{2}\left[ 2\left( Q_{1}+\tau \right) %
\right] ,  \label{PEI-13b}
\end{gather}%
Consequently, we can obtain an exact solution for the Hamiltonian (\ref%
{1.158}), by solving the Hamilton equations for $Q_{1},Q_{2},P_{1}$ and $%
P_{2}$ as a function of $\tau $ and substituting them in Eq. (\ref{PEI-10}).
Nevertheless, we recall that it is not our goal. Rather, we shall obtain a
set of approximate equations of motion with easy solution in terms of
elementary functions. For this, we first consider in Eq. (\ref{PEI-10}),
that $Q_{i}$ and $P_{i}$ are functions that vary very slowly within a period 
$\pi $ of the $\theta $ coordinate\footnote{%
This is a reasonable ansatz since those variables are constant when we use
the non-perturbed Hamiltonian $H_{0}$. The time evolution of these variables
arises from the introduction of the (much smaller) Hamiltonian $H_{1}$.} (as
it is customary in the quasi-harmonic approximation). Hence, an approximate
solution for the canonical variables of the complete Hamiltonian (\ref{1.158}%
)\ is still given by (\ref{PEI-10}) but with $Q_{i}$ and $P_{i}$
representing solutions of the Hamilton equations with $K$ replaced by its
average over a period $T$, this approximation yields 
\end{subequations}
\begin{subequations}
\label{PEI-14}
\begin{gather}
\left\langle K\right\rangle =\frac{1}{T}\int_{0}^{T}K\left( Q,P,t\right) dt=%
\frac{1}{16}\left( -3k^{2}+2P_{1}^{2}\right) ,  \label{PEI-14a} \\
\left\langle K\right\rangle =\frac{1}{16}\left( 3P_{2}^{2}-P_{1}^{2}\right) 
\label{PEI-14b}
\end{gather}%
where we have neglected the variations of all $Q_{i}$ and $P_{i}$ within a
period, and $k$ is given by (\ref{PEI-11}). Note that the new coordinates
are all cyclic in this averaged Hamiltonian. Consequently, under this
approximation all new canonical momenta are kept constant. The equations of
motion for these new coordinates become 
\end{subequations}
\begin{subequations}
\label{1.210}
\begin{gather}
\dot{P}_{1}=-\frac{\partial \left\langle K\right\rangle }{\partial Q_{1}}=0\
,\ \ \dot{P}_{2}=-\frac{\partial \left\langle K\right\rangle }{\partial Q_{2}%
}=0  \label{1.210a} \\
\dot{Q}_{1}=\frac{\partial \left\langle K\right\rangle }{\partial P_{1}}=-%
\frac{P_{1}}{8}\ ,\ \ \dot{Q}_{2}=\frac{\partial \left\langle K\right\rangle 
}{\partial P_{2}}=\frac{3P_{2}}{8}.  \label{1.210b}
\end{gather}%
The integration of these equations is straightforward 
\end{subequations}
\begin{subequations}
\label{1.211}
\begin{gather}
P_{1}=\alpha _{1}=H_{0}\ ,\ \ P_{2}=\alpha _{2}=P_{\varphi },  \label{1.211a}
\\
Q_{1}=-\frac{P_{1}}{8}\tau +Q_{10}\ ,\ \ Q_{2}=\frac{3P_{2}}{8}\tau +Q_{20},
\label{1.211b}
\end{gather}%
the constants $Q_{10}$ and $Q_{20}$ are obtained by using the initial
conditions and evaluating (\ref{PEI-10}) and (\ref{1.211b})\ at $\tau =0.$

\subsection{Approximate solution for the ideal spherical pendulum}

Let us find an approximate solution for the ideal spherical pendulum
associated with the elliptical mode characterized by initial conditions in
which the pendulum is released with a small initial amplitude $\theta _{0}$
and a small initial precession but without initial nutation (initial
conditions with respect to the earth). As we have discussed, owing to the
rotation of the earth the initial precession with respect to an inertial
reference frame is slightly different. Therefore, in an inertial reference
frame the initial conditions become 
\end{subequations}
\begin{subequations}
\label{1.163}
\begin{equation}
\theta \left( 0\right) =\theta _{0}\ ,\ \ \varphi \left( 0\right) =0,\ \text{%
\ }\dot{\theta}\left( 0\right) =0\ ,\ \ \dot{\varphi}\left( 0\right) =\dot{%
\varphi}_{0},  \label{1.163b}
\end{equation}%
we also take $l=1m$. According with Eq. (\ref{tc1}), the initial conditions (%
\ref{1.163b}) are transformed into 
\end{subequations}
\begin{equation}
\bar{\theta}\left( 0\right) =\sin \theta _{0}\ ,\ \ \bar{P}_{\theta }\left(
0\right) =P_{\theta }(0)\sec \theta _{0}  \label{1.163c}
\end{equation}%
By applying the initial conditions (\ref{1.163b}) in Eqs. (\ref{1.154}, \ref%
{1.157}) we see that $P_{\theta }(0)=0$. Moreover, by combining Eqs. (\ref%
{alfa1}, \ref{PEI-7}, \ref{1.211a}, \ref{1.211b}), we obtain the following
expressions for the constants 
\begin{subequations}
\label{3.216}
\begin{gather}
\alpha _{2}=\frac{\dot{\varphi}_{0}}{\omega _{0}}\sin ^{2}\theta _{0}=\frac{%
\dot{\varphi}_{0}}{\omega _{0}}\bar{\theta}_{0}^{2}\ ,\ \ \alpha _{1}=\frac{%
\text{\ }\bar{\theta}_{0}^{2}}{2}+\frac{\alpha _{2}^{2}}{2\text{\ }\bar{%
\theta}_{0}^{2}}  \label{3.216a} \\
\ k=\frac{\text{\ }\bar{\theta}_{0}^{2}}{2}-\frac{\alpha _{2}^{2}}{2\text{\ }%
\bar{\theta}_{0}^{2}}  \label{3.216b} \\
Q_{10}=\frac{\pi }{4}\ ,\ \ Q_{20}=-\arctan \left( \frac{\bar{\theta}_{0}^{2}%
}{\alpha _{2}}\right) ,  \label{3.216c}
\end{gather}%
The new coordinates (\ref{1.211b}) are 
\end{subequations}
\begin{equation}
Q_{1}=-\frac{\alpha _{1}}{8}\tau +\frac{\pi }{4}\ ,\ \ Q_{2}=\frac{3P_{2}}{8}%
\tau -\arctan \left( \frac{\bar{\theta}_{0}^{2}}{\alpha _{2}}\right) ,
\end{equation}%
and the approximate solutions (\ref{PEI-10a}) and (\ref{PEI-10c}) yield 
\begin{subequations}
\label{3.212}
\begin{gather}
\bar{\theta}=\sqrt{\alpha _{1}+k\cos \left[ 2\left( 1-\frac{\alpha _{1}}{8}%
\right) \tau \right] },  \label{3.212a} \\
\varphi =Q_{2}+\arctan \left[ \frac{1}{\alpha _{2}}\left( k+\alpha _{1}\tan
\left( Q_{1}+\tau \right) \right) \right] ,  \label{3.212b}
\end{gather}%
In Figure \ref{Cap1Fig11} we show the solutions obtained with the
approximation (\ref{3.212}) and the exact solutions for the initial
conditions $\theta _{0}=0.1$rad, $\dot{\varphi}_{0}=1.00167$ rad/s. In part $%
A$ of this figure we superpose the graphics of the polar angle $\bar{\theta}$
and $\sin \theta $. Note that they cannot be distinguished. In part $B$ it
is shown the exact (increasing) azimuthal angle\ and the approximate one
(monotonic piecewise). It is observed that both graphics coincide for $0\leq
\tau \leq \tau _{\pi /2}$ where $\tau _{\pi /2}$ corresponds to the value of 
$\tau $ for which the phase of the function $\tan x$ in (\ref{3.212b}) is
equal to $\pi /2$. It is clear that the discontinuities in the derivative
appear for values $\left( 2n+1\right) \pi /2$ of the argument where $%
n=1,2,\ldots $%
\begin{figure}[tbh]
\begin{center}
\includegraphics[scale=1.5]{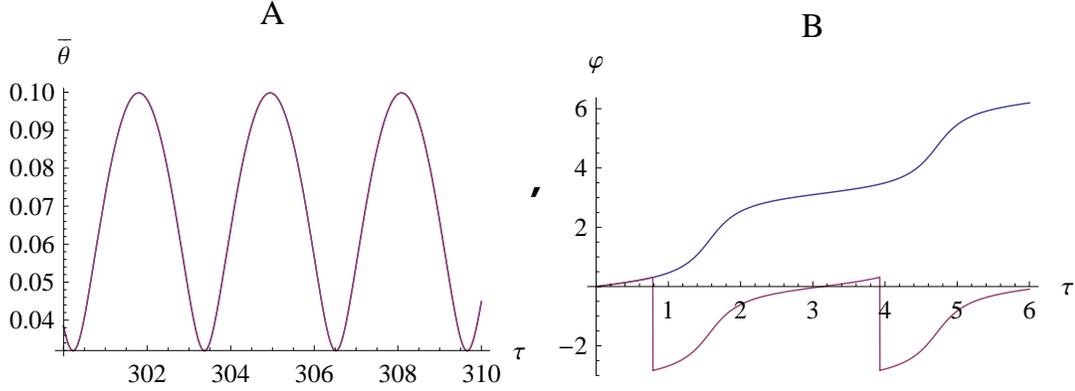}
\end{center}
\caption{\emph{(A) Graphical form of the approximate solution for }$\bar{%
\protect\theta}\ $\emph{with\ 300~}$\leq \protect\tau \leq 310$\emph{. The\
graphics of the exact solution cannot be distinguished from the approximate
one. (B)\ Graphical form of the (monotonic piecewise)\ approximate solution
for the angle }$\protect\varphi $\emph{, and of the smooth approximate
solution after the introduction of the }$\protect\gamma \ $\emph{angle, for }%
$0\leq \protect\tau \leq 6$\emph{. The approximate (smooth) and exact
solutions are superposed. }}
\label{Cap1Fig11}
\end{figure}

However, we can correct such a problem by introducing the angle $\gamma $
such that 
\end{subequations}
\begin{equation}
\arctan \left[ \frac{k+\alpha _{1}\tan \left( Q_{1}+\tau \right) }{\alpha
_{2}}\right] =\left( Q_{1}+\tau \right) +\arctan \gamma ,  \label{iden}
\end{equation}%
solving for $\gamma $ we obtain%
\begin{equation}
\gamma =\frac{2k\cos ^{2}\left( Q_{1}+\tau \right) +(\alpha _{1}-\alpha
_{2})\sin \left( 2\left( Q_{1}+\tau \right) \right) }{(\alpha _{1}+\alpha
_{2})-(\alpha _{1}-\alpha _{2})\cos \left( 2\left( Q_{1}+\tau \right)
\right) +k\sin \left( 2\left( Q_{1}+\tau \right) \right) }.  \label{3.213}
\end{equation}%
Note that by means of the identity (\ref{iden}), the strictly increasing
phase $\left( Q_{1}+\tau \right) $ has been extracted from the argument of
the function $\tan x,$ and we have introduced a phase $\gamma $ that
oscillates and remains always finite as argument of the function $\arctan x$%
. Substituting in (\ref{3.212b}) we finally get 
\begin{equation}
\varphi =Q_{2}+\left( Q_{1}+\tau \right) +\arctan \gamma ,  \label{3.215}
\end{equation}%
and the parameter $Q_{20}$ in Eq. (\ref{3.216c}) can be rewritten in terms
of the initial conditions as%
\begin{equation}
Q_{20}=-\frac{\pi }{4}+\arctan \left( \frac{\alpha _{2}-\bar{\theta}_{0}^{2}%
}{\bar{\theta}_{0}^{2}+\alpha _{2}}\right)
\end{equation}%
Figure \ref{Cap1Fig12} shows the superposition of the graphics for the
azimuthal angle $\varphi $ (curve) obtained from the exact solution and the
approximation (\ref{3.215}), for $500\leq \tau \leq 510$.\ Like in the case
of the polar angle we observe that the exact and approximate solutions
cannot be distinguished from each other. The interval to plot was chosen in
order to exhibit the asymptotic behavior of the approximate solution.

\begin{figure}[tbh]
\begin{center}
\includegraphics[scale=1]{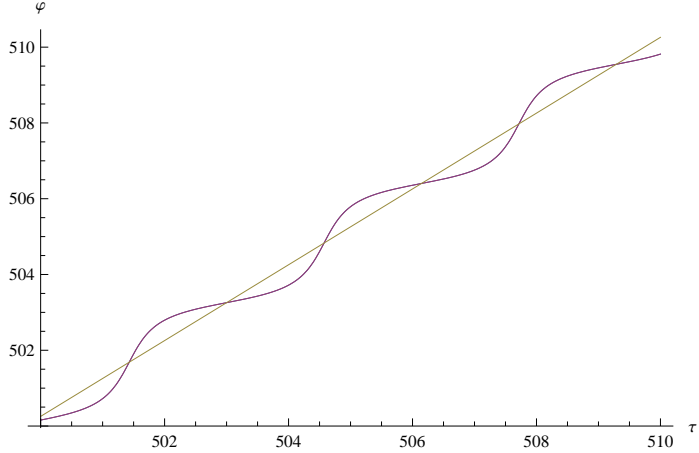}
\end{center}
\caption{\emph{Graphical form of the approximate values of the angle $%
\protect\varphi $ for $500\leq \protect\tau \leq 510$. The straight line
corresponds to the linear approximation given by (\protect\ref{al}). The
approximate and exact solutions are superposed.}}
\label{Cap1Fig12}
\end{figure}

The period of motion is by definition twice the period of the coordinate $%
\bar{\theta}$ (\ref{3.212a}), such that in this approximation the period (in
dimensionless units) is given by%
\begin{equation}
T=\frac{4\pi }{2\left( 1-\frac{\alpha _{1}}{8}\right) }=\frac{16\pi }{%
8-\alpha _{1}},  \label{ppie}
\end{equation}%
and for our current example it yields $T=2.00847$s, whose deviation with
respect to the exact period is one part of $10^{6}$ \cite{Brizard}. From Eq.
(\ref{3.212b}) it is followed that the approximate solution for $\dot{\varphi%
}$ has the same period $T$ of $\theta $. The straight line shown in Fig. \ref%
{Cap1Fig12} corresponds to the linear approximation for the function $%
\varphi \left( \tau \right) $ given by%
\begin{equation}
\tilde{\varphi}_{a}(\tau )\equiv \varphi _{0}+\frac{\varphi \left( T\right)
-\varphi \left( 0\right) }{T}\tau =\frac{1}{8}\left( 8-\alpha _{1}+3\alpha
_{2}\right) \tau .  \label{al}
\end{equation}%
Finally, we obtain the average angular velocity of the apsidal precession.
That is, the quotient between the angular excess over $2\pi $ and the period
of the motion%
\begin{equation}
\omega _{a}=\frac{\varphi \left( T\right) -\varphi \left( 0\right) -2\pi }{T}%
=\frac{3}{8}\alpha _{2},  \label{wa}
\end{equation}%
and using Eq. (\ref{3.216a}) we see that it coincides with the angular
velocity in the Airy's apsidal precession \cite{Airy} 
\begin{equation}
\omega _{a}=\frac{3}{8}\alpha _{2}=\frac{3}{8}\frac{\dot{\varphi}_{0}}{%
\omega _{0}}\sin ^{2}\theta _{0}.
\end{equation}

\subsubsection{Geometrical interpretation of the solution}

\begin{figure}[tbh]
\begin{center}
\includegraphics[scale=1]{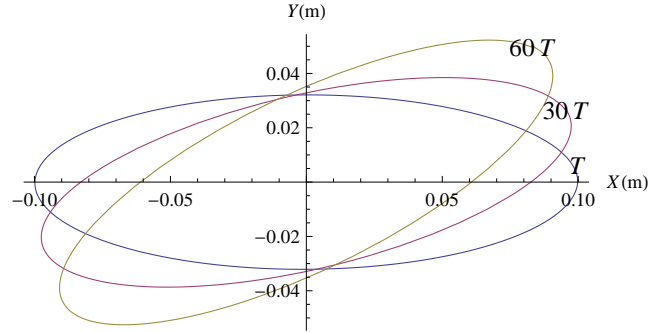}
\end{center}
\caption{\emph{Projection of the trajectory in the $XY\ plane$ for $30nT\leq
t\leq (30n+1)T$ \ with $n=0,1,2,$ using the approximate solution (\protect
\ref{1.213}). The approximate and exact solutions cannot be distinguished.}}
\label{Cap1Fig13}
\end{figure}

We conclude the study of the dynamics of the ISP by expressing the solutions
obtained so far in terms of the cartesian coordinates. As we shall see, it
permits us to give a simple geometrical interpretation to the projection of
the motion on the $XY\ $plane, when we use the approximate solutions (\ref%
{3.212}). In terms of the spherical coordinates, the dimensionless cartesian
coordinates yield%
\begin{equation}
q_{1}\equiv \frac{x}{l}=\sin \theta \cos \varphi \ ,\ \ q_{2}\equiv \frac{y}{%
l}=\sin \theta \sin \varphi ,
\end{equation}%
and replacing the approximate solutions for $\theta $ and $\varphi $, we
obtain an approximate solution for the dimensionless cartesian coordinates
(see appendix)%
\begin{equation}
\left[ 
\begin{array}{c}
q_{1} \\ 
q_{2}%
\end{array}%
\right] =%
\begin{bmatrix}
\cos \left( \omega _{a}\tau \right)  & -\sin \left( \omega _{a}\tau \right) 
\\ 
\sin \left( \omega _{a}\tau \right)  & \cos \left( \omega _{a}\tau \right) 
\end{bmatrix}%
\begin{bmatrix}
\bar{\theta}_{0}\cos \left( \omega \tau \right)  \\ 
\left( \alpha _{2}/\bar{\theta}_{0}\right) \sin \left( \omega \tau \right) 
\end{bmatrix}
\label{1.213}
\end{equation}%
Figure \ref{Cap1Fig13} shows the projection of the trajectory on the $XY$
plane for the initial conditions $\theta _{0}=0.1$ rad, $\dot{\varphi}%
_{0}=1.00167~Rad/s$ and $30nT\leq t\leq (30n+1)T$ with $n=0,\ 1,$ $2$; using
the approximation (\ref{1.213}) and the exact solution. Once again, the
comparison shows that the superposition of both solutions makes them
indistinguishable.

It is easy to see that the parametric equations (\ref{1.213}) describe an
ellipse of semiaxes $a$ and$\ b$ with angular frequency $\omega $ (in
standard units), given by%
\begin{gather}
a=l\bar{\theta}_{0}=l\sin \theta _{0},\ b=\frac{l\alpha _{2}}{\bar{\theta}%
_{0}},  \label{abdef} \\
\omega =\omega _{0}\left( 1-\frac{\alpha _{1}}{8}\right) ,\ \alpha _{1}=%
\frac{\text{\ }\bar{\theta}_{0}^{2}}{2}+\frac{\alpha _{2}^{2}}{2\text{\ }%
\bar{\theta}_{0}^{2}},  \label{omegaalfa1}
\end{gather}%
the matrix of rotation in Eq. (\ref{1.213}) shows that such an ellipse
precesses counterclockwise with angular frequency%
\begin{equation}
\omega _{a}=\left( \frac{3}{8}\alpha _{2}\right) \omega _{0}=\frac{3}{8}%
\frac{ab}{l^{2}}\omega _{0},
\end{equation}%
the last expression is the so-called apsidal Airy's precession \cite{Airy}.
As a matter of consistency, when $\dot{\varphi}_{0}=0,$ and $\omega _{a}=0,$
the difference of phase $\alpha _{1}/8$ given by Eq. (\ref{omegaalfa1})
coincides with the relative variation of first order of the frequency of a
plane pendulum with finite amplitude $\theta _{0}$ \cite{Goldstein}. When $%
\dot{\varphi}_{0}\neq 0$, equation (\ref{3.216a}) shows that $\alpha
_{2}\neq 0$, so that this variation is corrected by $\alpha _{2}^{2}/(16$\ $%
\bar{\theta}_{0}^{2})$ as an effect of the precession. Finally, we should
emphasize that the expression obtained by us for the apsidal Airy's
precession does not depend on the a priori assumption that Airy makes with
respect to the elliptical trajectories of the projection of the pendulum
motion \cite{Airy}. Such a feature is deduced in our framework in a natural
manner as a consequence of the slow variation of the new coordinates. We
point out that the method also provides the correction to the period up to
second order in the coordinates.

\section{Physical Symmetrical Pendulum\label{sec:PSP}}

\begin{figure}[tbh]
\begin{center}
\includegraphics[scale=0.8]{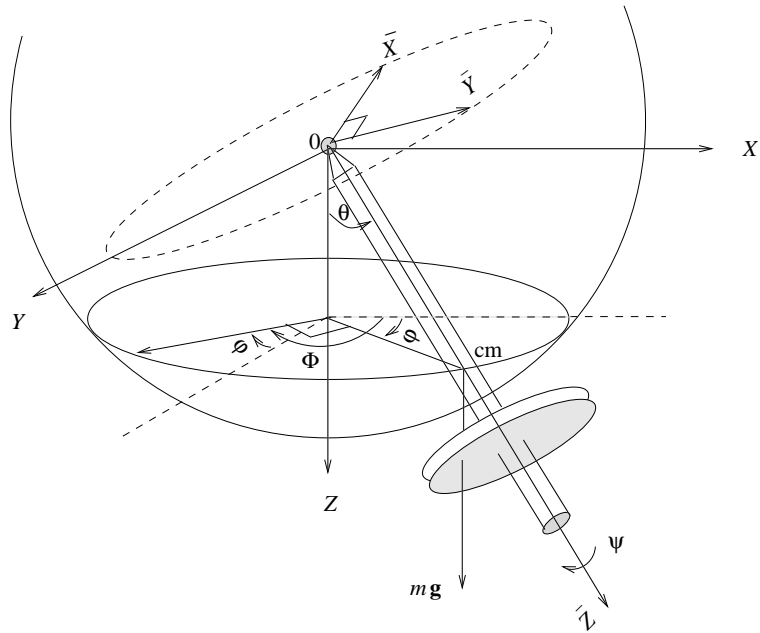}
\end{center}
\caption{\emph{Physical Symetrical Pendulum hung on an edge. $\protect\theta %
,\ \Phi $ and $\protect\psi $ are the Euler angles used in the description
of the motion. The spherical coordinate $\protect\varphi $ is shown as well
as its relation with the Euler angle$\ \Phi .$ }}
\label{Cap3Fig1}
\end{figure}

We define a Physical Symmetrical Pendulum (PSP) as a rigid solid with an
axis of symmetry that can rotate freely around a fixed point (support)
located at one edge of the symmetry axis and in which the center of mass
(CM) is located below the fixed point. An example is a disc supported by a
cylindrical rod hung on an edge, as shown in Fig. \ref{Cap3Fig1}. A
comparison with the symmetrical top shows us that indeed we are dealing with
the same system, and the only difference is the range in which nutation
motion occurs: $0\leq \theta \leq \pi /2$ for the PSP, and $\pi /2\leq
\theta \leq \pi $ for the top\footnote{%
We are using the convention of positive direction of the $Z-$axis in the
direction of the gravitational field (i.e. downwards).}.

We use the Euler angles $\theta ,\ \Phi ,\ \psi \ $as\ generalized\
coordinates. As shown in Fig. \ref{Cap3Fig1}, these angles provide the
orientation of the system of axes $\bar{X}\bar{Y}\bar{Z}$ fixed to the
pendulum, with respect to the inertial system of axes $XYZ$. We can note
that there is a difference of phase of $\pi /2\ $between the Euler angle $%
\Phi $ and the azimuthal angle $\varphi $ of the spherical coordinates, that
is%
\begin{equation}
\Phi =\varphi +\pi /2.  \label{3.14}
\end{equation}%
The components of the angular velocity $\boldsymbol{\omega }$ in the basis
of axes $\bar{X}\bar{Y}\bar{Z}$ fixed to the pendulum, yield \cite{Goldstein}%
\begin{eqnarray}
\omega _{\bar{x}} &=&\dot{\Phi}\sin \theta \sin \psi +\dot{\theta}\cos {\psi
,}  \notag \\
\omega _{\bar{y}} &=&\dot{\Phi}\sin \theta \cos \psi -\dot{\theta}\sin \psi ,
\notag \\
\omega _{\bar{z}} &=&\dot{\Phi}\cos \theta +\dot{\psi}  \label{3.7}
\end{eqnarray}%
it is clear that in the system of axis $\bar{X}\bar{Y}\bar{Z}$, the position 
$\mathbf{r}_{\mathbf{cm}}$ of the CM, gives 
\begin{equation}
\mathbf{r}_{\mathbf{cm}}=l\mathbf{\bar{k}}
\end{equation}%
where $l$ is the distance from the fixed point to the CM, while in the
inertial system of axis $XYZ$, it becomes%
\begin{equation}
\mathbf{r}_{\mathbf{cm}}=l\sin \theta \sin \Phi \ \mathbf{i}\ -l\sin \theta
\cos \Phi \ \mathbf{j}+l\cos \theta \ \mathbf{k}.  \label{3.13}
\end{equation}%
Finally, the Lagrangian of the system is given by%
\begin{equation}
L=\tfrac{1}{2}\left[ I_{\bar{x}}\left( \dot{\theta}^{2}+\dot{\Phi}^{2}\sin
^{2}\theta \right) +I_{\bar{z}}\left( \dot{\Phi}\cos \theta +\dot{\psi}%
\right) ^{2}\right] +mgl\cos \theta ,  \label{L}
\end{equation}%
where $I_{\bar{x}}$ and $I_{\bar{z}}$ are the moments of inertia of the
pendulum with respect to the axes $\bar{X}$ and $\bar{Z}$ fixed to the body.
The constant canonical momenta associated with the cyclic coordinates $\Phi $
and $\psi $ are given by%
\begin{eqnarray}
p_{\Phi } &=&\frac{\partial L}{\partial \dot{\Phi}}=I_{\bar{x}}\dot{\Phi}%
\sin ^{2}\theta +I_{\bar{z}}\cos \theta ~\left( \dot{\psi}+\dot{\Phi}\cos
\theta \right)  \label{pfipsi PSP1} \\
p_{\psi } &=&\frac{\partial L}{\partial \dot{\psi}}=I_{\bar{z}}\left( \dot{%
\Phi}\cos \theta +\dot{\psi}\right)  \label{pfipsi PSP2}
\end{eqnarray}

\subsection{Hamiltonian of the PSP}

The Lagrangian (\ref{L}) is a homogenous function of second degree, the
transformation from cartesian to generalized coordinates does not depend on
time, and the potential does not depend on the generalized velocities. Thus,
the Hamiltonian $h$ becomes the total energy of the system and can be
expressed in canonical variables as follows 
\begin{equation}
h=\frac{1}{2}\left( \frac{p_{\theta }^{2}}{I_{\bar{x}}}+\frac{\left( p_{\psi
}\cos \theta -p_{\Phi }\right) ^{2}}{I_{\bar{x}}\sin ^{2}\theta }+\frac{%
p_{\psi }^{2}}{I_{\bar{z}}}\right) -mgl\cos \theta ,  \label{3.107}
\end{equation}%
or in dimensionless units 
\begin{equation}
H=\frac{1}{2}\left( P_{\theta }^{2}+\frac{\left( P_{\Phi }-P_{\psi }\cos
\theta \right) ^{2}}{\sin ^{2}\theta }+\alpha P_{\psi }^{2}\right) -\cos
\theta ,  \label{3.113}
\end{equation}%
where we have used the definitions 
\begin{equation}
\Omega _{0}\equiv \sqrt{\frac{mgl}{I_{\bar{x}}}}\ ,\ \ \tau \equiv \Omega
_{0}t\ ,\ \ H\equiv \frac{h}{mgl}\ ,\ \ P_{\psi ,\Phi }\equiv \frac{p_{\psi
,\Phi }}{\sqrt{mglI_{\bar{x}}}}\ ,\ \ \alpha \equiv \frac{I_{\bar{x}}}{I_{%
\bar{z}}}  \label{dimensionless PSP}
\end{equation}%
the parameter $\alpha \ $accounts on the shape of the pendulum. For a PSP
like the one represented in Figure \ref{Cap3Fig1}, $\alpha >1$ and its
ellipsoid of inertia is prolate. In this study we are interested in this
case. When $\alpha <1$ we would have an oblate ellipsoid of inertia while $%
\alpha =1$ corresponds to an spherical ellipsoid.

Now we proceed in a way similar to the case of the ISP. That is, we
introduce a generating function similar to (\ref{tcp}), that leaves the
variables of precession and spin unaltered%
\begin{equation}
F_{2}(\theta ,\Phi ,\psi )=\bar{P}_{\theta }\sin \theta +\bar{P}_{\Phi }\Phi
+\bar{P}_{\psi }\psi ,  \label{3.114}
\end{equation}%
which leads to the following transformation between canonical coordinates%
\begin{equation}
\bar{\theta}=\sin \theta \ ,\ \ P_{\theta }=\bar{P}_{\theta }\cos \theta ,
\label{3.121}
\end{equation}%
and the Hamiltonian becomes%
\begin{equation}
\bar{H}=\frac{1}{2}\left( \bar{P}_{\theta }^{2}\left( 1-\bar{\theta}%
^{2}\right) +\frac{\left( P_{\Phi }-P_{\psi }\sqrt{1-\bar{\theta}^{2}}%
\right) ^{2}}{\bar{\theta}^{2}}+\alpha P_{\psi }^{2}\right) -\sqrt{1-\bar{%
\theta}^{2}},
\end{equation}%
where we have preserved the symbols of the variables that describe
precession and spin. Expanding up to fourth order in $\bar{\theta}$ and
neglecting constant terms, the Hamiltonian becomes%
\begin{equation}
\bar{H}=\frac{1}{2}\left( \bar{P}_{\theta }^{2}\left( 1-\bar{\theta}%
^{2}\right) +\frac{\left[ P_{\Phi }-P_{\psi }\left( 1-\frac{\bar{\theta}^{2}%
}{2}\right) \right] ^{2}}{2\bar{\theta}^{2}}+\alpha P_{\psi }^{2}\right) +%
\frac{\bar{\theta}^{2}}{2}+\frac{\bar{\theta}^{4}}{8},  \label{HB}
\end{equation}%
for future purposes, it is convenient the following separation%
\begin{eqnarray}
\bar{H} &=&\bar{H}_{0}+\bar{H}_{1}  \label{3.108ab} \\
\bar{H}_{0} &\equiv &\frac{\bar{P}_{\theta }^{2}}{2}+\frac{\bar{\theta}^{2}}{%
2}+\frac{\left( P_{\Phi }-P_{\psi }\right) ^{2}}{2\bar{\theta}^{2}}-\frac{%
\bar{P}_{\theta }^{2}\bar{\theta}^{2}}{2}+\frac{\bar{\theta}^{4}}{8},
\label{3.108a} \\
\bar{H}_{1} &\equiv &\frac{P_{\psi }}{2}\left[ P_{\Phi }+P_{\psi }(\alpha -1)%
\right] +\frac{P_{\psi }^{2}}{8}\bar{\theta}^{2},  \label{3.108b}
\end{eqnarray}
It is easy to realize that this separation is not consequent with the
perturbation theory, since in both $\bar{H}_{0}$ and $\bar{H}_{1}$ there are
terms of order two and four. Indeed this separation has the aim of matching $%
\bar{H}_{0}$ with the Hamiltonian of the ISP (\ref{1.158}), instead of
implementing a standard perturbation theory.

To do that, we introduce a new generating function 
\begin{equation}
G_{2}\left( \Phi ,\psi ,\bar{P}_{\Phi },\bar{P}_{\psi }\right) =\Phi (\bar{P}%
_{\Phi }+\bar{P}_{\psi })+\psi \bar{P}_{\psi },
\end{equation}%
which leaves invariant the variables that describe the nutation and
transforms the precession and spin variables as follows 
\begin{subequations}
\begin{gather}
\bar{\Phi}=\frac{\partial G_{2}}{\partial \bar{P}_{\Phi }}=\Phi \ ,\ \
P_{\Phi }=\frac{\partial G_{2}}{\partial \Phi }=\bar{P}_{\Phi }+\bar{P}%
_{\psi }  \label{3.120a} \\
\bar{\psi}=\frac{\partial G_{2}}{\partial \bar{P}_{\psi }}=\psi +\Phi \ ,\ \
P_{\psi }=\frac{\partial G_{2}}{\partial \psi }=\bar{P}_{\psi }.
\label{3.120b}
\end{gather}%
and the new Hamiltonian becomes 
\end{subequations}
\begin{subequations}
\begin{gather}
\bar{H}_{0}=\frac{\bar{P}_{\theta }^{2}}{2}+\frac{\bar{\theta}^{2}}{2}+\frac{%
\bar{P}_{\Phi }^{2}}{2\bar{\theta}^{2}}-\frac{\bar{P}_{\theta }^{2}\bar{%
\theta}^{2}}{2}+\frac{\bar{\theta}^{4}}{8},  \label{3.115a} \\
\bar{H}_{1}=\frac{\bar{P}_{\psi }}{2}(\bar{P}_{\Phi }+\alpha \bar{P}_{\psi
})+\frac{\bar{P}_{\psi }^{2}}{8}\bar{\theta}^{2}.  \label{3.115b}
\end{gather}%
Since (\ref{3.115a}) coincides with (\ref{1.158}), we can consider equations
(\ref{PEI-10}) as the solution of the Hamiltonian $\bar{H}_{0}$ and then we
construct an additional CT that permits to reduce the new Hamiltonian to $%
\bar{H}_{1}$ alone. This is the task of the next section.

\subsection{Variation of Parameters}

According with the results of the previous section, we consider the
approximate solution (\ref{PEI-10}) as the solution of (\ref{3.115a}), that
is 
\end{subequations}
\begin{subequations}
\label{3.116}
\begin{gather}
\bar{\theta}=\sqrt{P_{1}+k\sin \left[ 2\left( Q_{1}+\tau \right) \right] },
\label{3.116a} \\
\bar{P}_{\theta }=\frac{k\cos \left[ 2\left( Q_{1}+\tau \right) \right] }{%
\bar{\theta}}\ ,\ \ \bar{P}_{\Phi }=P_{2},  \label{3.116b} \\
\bar{\Phi}=Q_{2}+\arctan \left[ \frac{1}{P_{2}}\left( k+P_{1}\tan \left(
Q_{1}+\tau \right) \right) \right] ,  \label{3.116c}
\end{gather}%
with 
\end{subequations}
\begin{subequations}
\label{3.117}
\begin{gather}
Q_{1}=-\frac{P_{1}}{8}\tau +Q_{10}\ ,\ \ Q_{2}=\frac{3P_{2}}{8}\tau +Q_{20},
\label{3.117a} \\
P_{1}=\frac{\bar{P}_{\theta 0}^{2}}{2}+\frac{\bar{\theta}_{0}^{2}}{2}+\frac{%
\bar{P}_{\Phi 0}^{2}}{2\bar{\theta}_{0}^{2}},  \label{3.117b} \\
\ P_{2}=\bar{P}_{\Phi }=P_{\Phi }-P_{\psi },  \label{3.117c} \\
\ k^{2}=P_{1}^{2}-P_{2}^{2}.  \label{3.117d}
\end{gather}

This solution suggests to construct a last CT to the new variables $\Gamma
_{i},\ \Pi _{i},$ with $i=1,2,$ replacing the constants $Q_{i0}$ by $\Gamma
_{i},\ $and $P_{i}$ by $\Pi _{i}$ (variation of parameters),$\ $while
leaving unaltered the variables that describe the spin degrees of freedom 
\end{subequations}
\begin{subequations}
\label{3.118}
\begin{gather}
Q_{1}=-\frac{\Pi _{1}}{8}\tau +\Gamma _{1}\ ,\ \ Q_{2}=\frac{3\Pi _{2}}{8}%
\tau +\Gamma _{2},  \label{3.118a} \\
Q_{3}=\Gamma _{3}=\bar{\psi},\ P_{1}=\Pi _{1}\ ,\ \ P_{2}=\Pi _{2},\ \Pi
_{3}=\bar{P}_{\psi }  \label{3.118b}
\end{gather}%
Indeed, the CT given by (\ref{3.118}) is a particular case of a more general
class of canonical transformations $(Q,\ P)\rightarrow \left( \Gamma ,\ \Pi
\right) $ that we call Linear Transformations (see appendix). There is a
generating function of type II associated with these transformation $S\left(
Q_{1},Q_{2},Q_{3},\Pi _{1},\Pi _{2},\Pi _{3},\tau \right) $, and the new
Hamiltonian $K$ as a function of $\Gamma _{i}$ and $\Pi _{i}$ is obtained
from 
\end{subequations}
\begin{equation}
K=\bar{H}_{0}+\bar{H}_{1}+\frac{\partial S}{\partial \tau }=\bar{H}%
_{1}\left( \Gamma ,\Pi \right) \ ,
\end{equation}%
where we have used our a priori assumption that (\ref{3.116}) is the exact
solution for$\ \bar{H}_{0},$ that is 
\begin{equation}
\bar{H}_{0}+\frac{\partial S}{\partial \tau }=0.
\end{equation}%
such that $\bar{H}_{1}$ described by (\ref{3.115b}) in terms of these new
variables, is the new Hamiltonian%
\begin{equation}
K=\bar{H}_{1}\left( \Gamma ,\Pi ,\right) =\frac{1}{2}\Pi _{3}\left( \Pi
_{2}+\alpha \Pi _{3}\right) +\frac{\Pi _{3}^{2}}{8}\left\{ \Pi _{1}+k\sin %
\left[ 2\left( \omega _{1}\tau +\Gamma _{1}\right) \right] \right\} ,
\end{equation}%
where $\omega _{1}=1-\Pi _{1}/8.$ It is clear that $\Gamma _{1}$ is still a
slowly varying coordinate, such that we can neglect its variation within a
period of time $T=2\pi /\omega _{1}$. Carrying out a \textquotedblleft fast
integration\textquotedblright\ with respect to $\tau $ we see that the
oscillating term vanishes and the averaged Hamiltonian becomes 
\begin{equation}
\left\langle K\right\rangle =\frac{1}{2}\Pi _{3}\left( \Pi _{2}+\alpha \Pi
_{3}\right) +\frac{1}{8}\Pi _{3}^{2}\Pi _{1},  \label{3.226}
\end{equation}%
it is observed that the coordinates $\Gamma $ are cyclic within this
averaged Hamiltonian so that the new momenta are constant (within our
approximation) and equal to their initial values%
\begin{equation}
\Pi _{1}=P_{1}\ ,\ \ \Pi _{2}=P_{2}\ ,\ \ \Pi _{3}=P_{\psi }\ 
\end{equation}%
while the equations of motion for the new coordinates are 
\begin{subequations}
\begin{gather}
\dot{\Gamma}_{1}=\frac{\partial \left\langle K\right\rangle }{\partial \Pi
_{1}}=\frac{1}{8}\Pi _{3}^{2}\ ,\ \ \dot{\Gamma}_{2}=\frac{\partial
\left\langle K\right\rangle }{\partial \Pi _{2}}=\frac{\Pi _{3}}{2}, \\
\dot{\Gamma}_{3}=\frac{\partial \left\langle K\right\rangle }{\partial \Pi
_{3}}=\frac{\Pi _{2}}{2}+\alpha \Pi _{3}+\frac{1}{4}\Pi _{3}\Pi _{1},
\end{gather}%
integrating out we have 
\end{subequations}
\begin{subequations}
\label{3.220}
\begin{gather}
\Gamma _{1}=\frac{1}{8}P_{\psi }^{2}\tau +Q_{10},  \label{3.220a} \\
\Gamma _{2}=\frac{P_{\psi }}{2}\tau +Q_{20},  \label{3.220b} \\
\Gamma _{3}=\left( \frac{P_{2}}{2}+\alpha P_{\psi }+\frac{1}{4}P_{\psi
}P_{1}\right) \ \tau +\bar{\psi}_{0}.  \label{3.220c}
\end{gather}

By virtue of these results, the relations (\ref{3.116}) are still
approximate solutions for the variables $\bar{\theta},$ $\bar{P}_{\theta },\ 
\bar{\Phi}$ of the system described by (\ref{HB}), where $P_{1},P_{2},$ and $%
k$ are the constants defined by (\ref{3.117b})-(\ref{3.117d}), while the
time evolution of the functions $Q_{i}$ in (\ref{3.118}) yields 
\end{subequations}
\begin{subequations}
\label{3.119}
\begin{eqnarray}
Q_{1} &=&\frac{1}{8}\left( P_{\psi }^{2}-P_{1}\right) \tau +Q_{10},
\label{3.119a} \\
Q_{2} &=&\left( \frac{3P_{2}}{8}+\frac{P_{\psi }}{2}\right) \tau +Q_{20}.
\label{3.119b}
\end{eqnarray}%
As for the spin angle $\psi $, it can be obtained from (\ref{3.120b}) 
\end{subequations}
\begin{equation}
\psi =\bar{\psi}-\Phi =\Gamma _{3}-\Phi .  \label{3.120}
\end{equation}

It is important to point out that despite the multiple canonical
transformations carried out to reduce the exact Hamiltonian (\ref{3.113}) to
the approximate Hamiltonian (\ref{3.226}), the new coordinates $\left(
\Gamma _{1},\ \Gamma _{2},\ \Gamma _{3}\right) $ has an apparent physical
description with respect to the original coordinates $\left( \theta ,\ \Phi
,\ \psi \right) $, that is: the coordinate $\Gamma _{1}$ still describes the
nutation, the coordinate$\ \Gamma _{2}$ still describes the precession and $%
\Gamma _{3}$ describes the spin. As we shall see later, this fact simplifies
considerably the geometrical analysis of the solutions. Now, we shall obtain
the approximate solutions associated with the elliptic mode with this
method, and then we compare them with the exact solutions. We are interested
in the geometrical interpretation of these solutions and then verify the
globality of them (that is we intend to check how close are the approximate
solutions with respect to the exact solutions for long intervals of time).

\subsection{Approximate solution for the PSP}

The initial conditions that characterize the elliptical mode in an inertial
reference frame are given by 
\begin{subequations}
\begin{gather}
\theta \left( 0\right) =\theta _{0}\ ,\ \ \Phi \left( 0\right) =\pi /2\ ,\ \
\psi \left( 0\right) =0,  \notag \\
\dot{\theta}\left( 0\right) =0\ ,\ \ \dot{\Phi}\left( 0\right) =\dot{\Phi}%
_{0}\ ,\ \ \dot{\psi}\left( 0\right) =0,
\end{gather}%
for these conditions, we have again that $P_{\theta }\left( 0\right) =0,$
and in terms of the new coordinates they transform according with (\ref%
{3.121}) into 
\end{subequations}
\begin{equation}
\bar{\theta}_{0}=\sin \theta _{0}\ ,\ \ \bar{P}_{\theta 0}=P_{\theta
}(0)\sec \theta _{0}.
\end{equation}%
The dimensionless constants of motion $P_{\Phi }$ and $P_{\psi }$ are
determined from these conditions and equations (\ref{pfipsi PSP1}, \ref%
{pfipsi PSP2}, \ref{dimensionless PSP}) 
\begin{subequations}
\label{PFS-11}
\begin{gather}
P_{\Phi }=\frac{p_{\Phi }}{\sqrt{mglI_{\bar{x}}}}=\frac{\dot{\Phi}_{0}}{%
\sqrt{mglI_{\bar{x}}}}\left[ I_{\bar{x}}\bar{\theta}_{0}^{2}+I_{\bar{z}%
}\left( 1-\bar{\theta}_{0}^{2}\right) \right] ,  \label{PFS-11a} \\
P_{\psi }=\frac{p_{\psi }}{\sqrt{mglI_{\bar{x}}}}=\frac{I_{\bar{z}}}{\sqrt{%
mglI_{\bar{x}}}}\left( \dot{\Phi}_{0}\sqrt{1-\bar{\theta}_{0}^{2}}\right) ,
\label{PFS-11b}
\end{gather}%
\textbf{\ }and the constants $P_{1},~P_{2},~Q_{10},~Q_{20}$ are obtained
from (\ref{3.117}) 
\end{subequations}
\begin{subequations}
\label{3.221}
\begin{gather}
P_{1}=\frac{P_{2}^{2}}{2\bar{\theta}_{0}^{2}}+\frac{\bar{\theta}_{0}^{2}}{2}%
\ ,\ \ P_{2}=P_{\Phi }-P_{\psi },  \label{3.221a} \\
\ k=\frac{\bar{\theta}_{0}^{2}}{2}-\frac{P_{2}^{2}}{2\bar{\theta}_{0}^{2}}.
\label{3.221b}
\end{gather}%
\end{subequations}
\begin{equation}
Q_{10}=\frac{\pi }{4}\ ,\ \ Q_{20}=\frac{\pi }{2}-\arctan \left( \frac{%
P_{1}+k}{P_{2}}\right) ,  \label{3.222}
\end{equation}

The coordinate $\bar{\theta}$, is given by (\ref{3.116a})%
\begin{equation}
\bar{\theta}=\sqrt{P_{1}+k\cos \left[ \frac{1}{4}\left( 8-P_{1}+P_{\psi
}^{2}\right) \tau \right] },  \label{3.227}
\end{equation}%
while the azimuthal angle $\Phi $ given by (\ref{3.116c}) presents the same
discontinuities observed in (\ref{3.212b}), it is worked out in a similar
way of the previous case by introducing the auxiliary function%
\begin{equation}
\gamma =\frac{2k\cos ^{2}\left( Q_{1}+\tau \right) +(P_{1}-P_{2})\sin \left[
2\left( Q_{1}+\tau \right) \right] }{(P_{1}+P_{2})-(P_{1}-P_{2})\cos \left[
2\left( Q_{1}+\tau \right) \right] +k\sin \left[ 2\left( Q_{1}+\tau \right) %
\right] },
\end{equation}%
and the parameter $Q_{20}$ in (\ref{3.222})\ can be rewritten in terms of
the initial conditions as%
\begin{equation}
\ Q_{20}=\frac{\pi }{4}-\arctan \left( \frac{\bar{\theta}_{0}^{2}-P_{2}}{%
\bar{\theta}_{0}^{2}+P_{2}}\right) ,
\end{equation}%
such that the approximate solution for the azimuthal angle becomes%
\begin{equation}
\Phi =Q_{2}+\left( Q_{1}+\tau \right) +\arctan \gamma .  \label{3.225}
\end{equation}%
And the spin angle $\psi $ can be obtained from (\ref{3.120}).

The period of the system, that by definition is twice the period of the
angle $\bar{\theta}$, is given by%
\begin{equation}
T=\frac{16\pi }{8-P_{1}+P_{\psi }^{2}},  \label{3.228}
\end{equation}%
comparing with the corresponding period associated with the ISP Eq. (\ref%
{ppie}), we note that the angular momentum associated with the spin $P_{\psi
}$ of the PSP, introduces two corrections: \textbf{(1)\ }In $P_{1}$ by means
of the definition (\ref{3.221a}) of $P_{2}\ $[compare with Eqs. (\ref{1.211a}%
, \ref{3.216a})], and \textbf{(2)} In the quadratic correction of the
denominator of (\ref{3.228}).

From (\ref{3.116c}) and (\ref{3.120}) we infer that $\dot{\Phi}$ and $\dot{%
\psi}$ are periodic functions with the same period $T$ given by (\ref{3.228}%
). Therefore, we can obtain linear approximations for the angles $\Phi $ and 
$\psi $ by using the following definitions%
\begin{gather}
\tilde{\Phi}\left( \tau \right) \equiv \Phi _{0}+\left[ \frac{1}{T}%
\int_{0}^{T}\dot{\Phi}\left( s\right) ds\right] \tau =\Phi _{0}+\left[ \frac{%
\Phi \left( T\right) -\Phi \left( 0\right) }{T}\right] \tau ,  \notag \\
\tilde{\Phi}\left( \tau \right) =\frac{\pi }{2}+\frac{1}{8}\left(
8-P_{1}+3P_{2}+4P_{\psi }+P_{\psi }^{2}\right) \tau ,  \label{apl1}
\end{gather}%
\begin{gather}
\tilde{\psi}\left( \tau \right) \equiv \psi _{0}+\left[ \frac{1}{T}%
\int_{0}^{T}\dot{\psi}\left( s\right) ds\right] \tau =\psi _{0}+\left[ \frac{%
\psi \left( T\right) -\psi \left( 0\right) }{T}\right] \tau ,  \notag \\
\tilde{\psi}\left( \tau \right) =\frac{1}{8}\left( -8+P_{1}+P_{2}-4P_{\psi
}-P_{\psi }^{2}+2P_{1}P_{\psi }+8\alpha P_{\psi }^{2}\right) \tau .
\label{apl2}
\end{gather}%
The mean angular velocity of apsidal precession (that we call Allais'
precession), can be obtained analogously to the procedure in (\ref{wa}) 
\begin{eqnarray}
\Omega _{\text{Allais}} &\equiv &\frac{\Phi \left( T\right) -\Phi (0)-2\pi }{%
T}  \label{3.229b} \\
\Omega _{\text{Allais}} &=&\frac{3P_{2}}{8}+\frac{P_{\psi }}{2}=\frac{1}{8}%
\left( 3P_{\Phi }+P_{\psi }\right) ,  \label{3.229}
\end{eqnarray}%
where we have used Eqs. (\ref{3.225}, \ref{3.228}) and Eq. (\ref{3.221a}).
By comparing Eq. (\ref{3.229}) with Eq. (\ref{wa}) we see clearly the
contribution of the spin to the Allais' precession, through the canonical
momentum $P_{\psi }$ [compare also the expressions (\ref{1.211a}) and (\ref%
{3.221a}) of $P_{2}$ for the ISP and PSP respectively]. We also note that
the correction introduced by the spin is of the same order of the $Z-$%
component of the angular momentum $P_{2}$.

Now, introducing the numerical values 
\begin{eqnarray}
I_{\bar{x}} &=&I_{\bar{y}}=1Kg\cdot m^{2}\ ,\ \ I_{\bar{z}}=10^{-2}Kg\cdot
m^{2}\ ,\ \ l=1m\ ,\ \ m=1Kg  \notag \\
\theta _{0} &=&0.1,\ \Phi _{0}=\pi /2\ ,\ \ \psi _{0}=0,\text{ ( rad)} 
\notag \\
\dot{\theta}_{0} &=&0,\ \dot{\Phi}_{0}=1,\ \dot{\psi}_{0}=0,\text{ (rad/seg),%
}  \label{numerical input PSP}
\end{eqnarray}%
we obtain

\begin{subequations}
\begin{gather}
P_{1}=0.0054868\ ,\ \ P_{2}=0.00316787\ ,\ \ P_{\psi }=0.00317842  \notag \\
\ k=0.00447991\ ,\ \ Q_{10}=\frac{\pi }{4}\ ,\ \ Q_{20}=0.306786\ ,\ \ \bar{%
\psi}_{0}=\pi /2.  \label{numerical output PSP}
\end{gather}%
the period of motion reads 
\end{subequations}
\begin{equation}
T=\frac{16\pi }{8-P_{1}+P_{\psi }^{2}}\frac{1}{\Omega _{0}}=2.00846\text{ s},
\label{period PSP}
\end{equation}%
which is in excellent agreement with the exact solution \cite{Brizard}. The
apsidal Allais' precession in this approximation gives 
\begin{equation}
\Omega _{\text{Allais}}=0.00871254\text{ rad/s.}  \label{omega Allais}
\end{equation}

\begin{figure}[tbh]
\begin{center}
\includegraphics[scale=1]{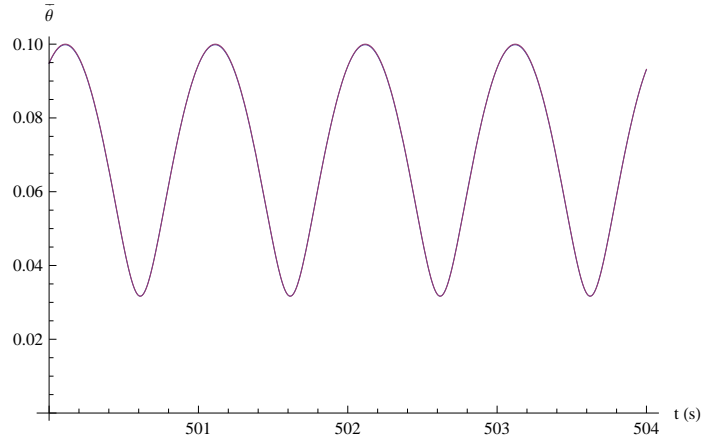}
\end{center}
\caption{\emph{Plot of the approximate solution (\protect\ref{3.227}) for
the nutation angle $\bar{\protect\theta}$ $\left( \sin \protect\theta %
\right) \ as\ a\ function\ of\ time,\ for$ $500\leq t\leq 504$ s. The exact
solution is superposed to the approximate one.}}
\label{Cap3Fig5}
\end{figure}

In Fig. \ref{Cap3Fig5} we compare the exact solution for the nutation angle $%
\theta $ with the approximate one Eq. (\ref{3.227}), for $500\leq t\leq 504$
s. It is observed that both solutions are superposed and cannot be
distinguished, showing the excellent agreement between them

\begin{figure}[tbh]
\begin{center}
\includegraphics[scale=1.6]{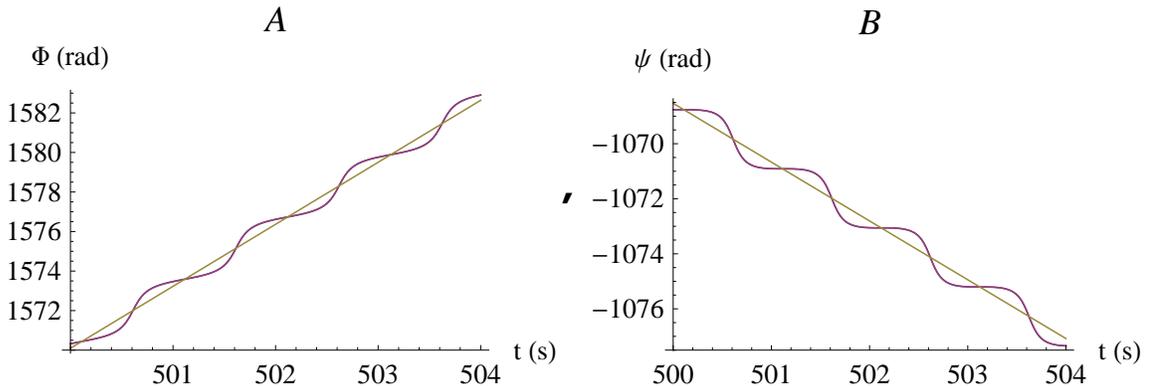}
\end{center}
\caption{\emph{(A) Approximate solutions for the azimuthal angle $\Phi ,$ for%
$\ $}$500\leq t\leq 504s$\emph{, (B) Approximate solutions of the spin angle 
$\protect\psi $, for$\ $}$500\leq t\leq 504s$\emph{. The straight lines
correspond to the linear approximations for each one of these angles. The
exact solutions cannot be distinguished from the approximate ones.}}
\label{Cap3Fig6}
\end{figure}

In Fig. \ref{Cap3Fig6} we compare the exact and approximate solutions for
the azimuthal angle $\Phi $ (\ref{3.225}) and the spin angle\ $\psi $ (\ref%
{3.120}), for $500\leq t\leq 504$ s. Once again, the solutions are
indistinguishable\ showing the excellent agreement between them. The
straight lines shown in these figures are the linear approximations (\ref%
{apl1}) and (\ref{apl2}). On the other hand, the interval of time chosen
shows the global character of the approximate solutions. 

\begin{figure}[tbh]
\begin{center}
\includegraphics[scale=0.8]{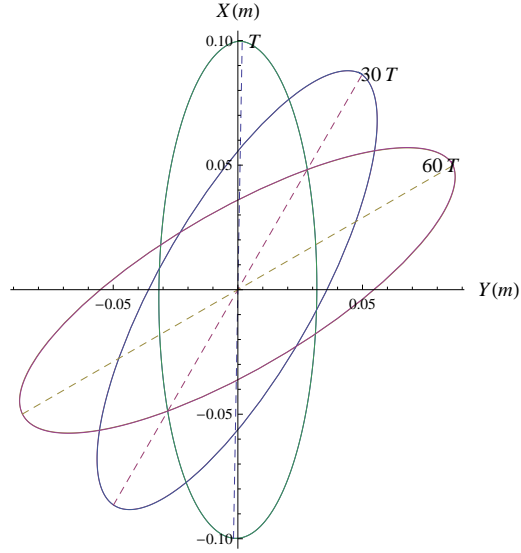}
\end{center}
\caption{\emph{Projections of the trajectory of the CM in the $XY-plane,$
obtained with the approximate solutions, for time intervals of a period $T,$
(A) for $0\leq t\leq T:(T).\ (B)\ For$ $29T\leq t\leq 30T:\ (30T).\ (C)$ for 
$59T\leq t\leq 60T:\ (60T).$ In all cases the exact solutions are superposed
to the approximate ones.}}
\label{Cap3Fig7}
\end{figure}

Finally, we compare in Fig. \ref{Cap3Fig7} the projections of the trajectory
in the $XY-$plane obtained from the exact solution and from the approximate
solution given by (see appendix):%
\begin{equation}
\left[ 
\begin{array}{c}
q_{1} \\ 
q_{2}%
\end{array}%
\right] =%
\begin{bmatrix}
\cos \left( \Omega _{\text{Allais}}\tau \right) & -\sin \left( \Omega _{%
\text{Allais}}\tau \right) \\ 
\sin \left( \Omega _{\text{Allais}}\tau \right) & \cos \left( \Omega _{\text{%
Allais}}\tau \right)%
\end{bmatrix}%
\begin{bmatrix}
\bar{\theta}_{0}\cos \left( \omega \tau \right) \\ 
\left( P_{2}/\bar{\theta}_{0}\right) \sin \left( \omega \tau \right)%
\end{bmatrix}
\label{2.213}
\end{equation}%
where%
\begin{gather}
q_{1}=\frac{x}{l}=\bar{\theta}\sin \Phi ,\ q_{2}=\frac{y}{l}=-\bar{\theta}%
\cos \Phi ,  \notag \\
\omega =\frac{1}{8}\left( 8-P_{1}+P_{\psi }^{2}\right) .  \label{q1q2PSP}
\end{gather}%
once again the solutions are practically identical. The dotted straight
lines show the evolution of the apsidal axis for the intervals of time
plotted, which obey the linear equation 
\begin{equation}
y=x\tan \left( \Omega _{\text{Allais}}\ nT\right) \ ,\ \ \text{with\ \ }%
n=1,\ 30,\ 60  \label{y=xtan(omeganT)}
\end{equation}

\section{Concluding remarks\label{sec:conclusions}}

We have obtained approximate solutions for the equations of motion of the
Ideal Spherical Pendulum (ISP) and the Physical Symmetrical Pendulum (PSP),
by using a Hamilton-Jacobi approach along with the technics of variation of
parameters and the averaging method. The success of the averaging method is
due to the presence of variables that vary very slowly within a period of
the polar angular coordinate.

In particular, we found approximate expressions for the apsidal precession
that undergoes an ISP and a PSP in their elliptical modes, when the pendula
are initiated with small initial amplitudes and small angular momenta of
precession with respect to an inertial reference frame. In addition, we
obtained approximate analytical expressions that describe the projections of
the trajectories on the horizontal plane for both types of pendula.

The method implemented in this paper to obtain these expressions is
operatively and conceptually simpler than the standard procedure used in
obtaining the exact expressions in terms of elliptical functions of Jacobi
and Weierstrass. In particular, our method permits to show with clarity the
origin of the precession of the ISP and of the PSP.\ Moreover, in the case
of the PSP, our approximate expressions introduce in a natural way the
correction due to the spin for its apsidal precession. Up to our knowledge,
this is a new approximation for an effect already observed in the
paraconical pendulum with symmetrical support.

On the other hand, this method also outlines the way to obtain approximate
solutions for the Physical Assymetrical Pendulum (paraconical pendulum),
which is still an open problem. It is because canonical transformations to
slowly varying coordinates are also possible for the Asymmetrical Pendulum,
and in terms of such kind of coordinates the new averaged Hamiltonian leads
to trivial (though approximate) equations of motion.

The examples presented here show an excellent numerical agreement of our
approximations with respect to the exact solutions. Further, such an
agreement is preserved even for long periods of time, so that they are
globally valid. Finally, as a consequence of the geometrical interpretation
of the approximate trajectories, we have associated the period of elliptical
motion $T\ $with the period of oscillations of the coordinate $\bar{\theta}$%
, this association permits to propose $T$ as the period of motion, which
coincides with the exact period up to the order of microseconds.

\section{Acknowledgments}

We acknowledge to División de Investigación de Bogotá (DIB) of Universidad
Nacional de Colombia, for its financial support.

\section{Appendix\label{sec:appendix}}

We present in this appendix the deduction of Eqs. (\ref{1.213}) and (\ref%
{2.213}), which expresses the geometrical interpretation that we have done
for the approximate solutions obtained in this paper. Further, we prove the
canonical character of the Linear Transformation introduced in (\ref{3.118}).

\begin{enumerate}
\item In order to deduce Eq. (\ref{1.213})
\end{enumerate}

\begin{equation}
\left[ 
\begin{array}{c}
q_{1} \\ 
q_{2}%
\end{array}%
\right] =%
\begin{bmatrix}
\cos \left( \omega _{a}\tau \right)  & -\sin \left( \omega _{a}\tau \right) 
\\ 
\sin \left( \omega _{a}\tau \right)  & \cos \left( \omega _{a}\tau \right) 
\end{bmatrix}%
\begin{bmatrix}
\bar{\theta}_{0}\cos \left( \omega \tau \right)  \\ 
\left( \alpha _{2}/\bar{\theta}_{0}\right) \sin \left( \omega \tau \right) 
\end{bmatrix}%
.  \label{E1}
\end{equation}%
we start with the following definitions 
\begin{subequations}
\label{E2}
\begin{gather}
q_{1}=\frac{x}{l}=\sin \theta \cos \varphi =\ \bar{\theta}\cos \varphi ,
\label{E2a} \\
q_{2}=\frac{y}{l}=\sin \theta \sin \varphi =\bar{\theta}\sin \varphi ,
\label{E2b}
\end{gather}%
and from the approximate solution for the angle of precession $\varphi \ $%
Eqs.$\ $(\ref{3.216})-(\ref{3.212b}) we get 
\end{subequations}
\begin{subequations}
\label{E4}
\begin{gather}
Q_{1}=-\frac{\alpha _{1}}{8}\tau +Q_{10}\ ,\ \ Q_{2}=\frac{3\alpha _{2}}{8}%
\tau +Q_{20},  \label{E4a} \\
\alpha _{1}=\frac{\text{\ }\bar{\theta}_{0}^{2}}{2}+\frac{\alpha _{2}^{2}}{2%
\text{\ }\bar{\theta}_{0}^{2}}\ ,\ \ k=\frac{\text{\ }\bar{\theta}_{0}^{2}}{2%
}-\frac{\alpha _{2}^{2}}{2\text{\ }\bar{\theta}_{0}^{2}},  \label{E4b} \\
k^{2}=\alpha _{1}^{2}-\alpha _{2}^{2}  \label{E4c} \\
Q_{10}=\frac{\pi }{4}\ ,\ \ Q_{20}=-\arctan \left( \frac{k+\alpha _{1}}{%
\alpha 2}\right)   \label{E4d}
\end{gather}%
\end{subequations}
\begin{gather}
\varphi =Q_{2}+\arctan \left\{ \frac{1}{\alpha _{2}}\left[ k+\alpha _{1}\tan
\left( \omega \tau +\frac{\pi }{4}\right) \right] \right\}   \label{E3a} \\
\varphi =\omega _{a}\tau +Q_{20}+\eta   \label{E3b}
\end{gather}%
where 
\begin{gather}
\eta \equiv \arctan \left\{ \frac{1}{\alpha _{2}}\left[ k+\alpha _{1}\tan
\left( \omega \tau +\frac{\pi }{4}\right) \right] \right\} ,  \label{E5a} \\
\omega _{a}=\frac{3\alpha _{2}}{8},\ \omega =1-\frac{\alpha _{1}}{8}.
\label{E5b}
\end{gather}%
From the definitions (\ref{E5a}) and (\ref{E4c}) it can be proved that 
\begin{subequations}
\label{E7}
\begin{gather}
\cos \eta =\frac{\alpha _{2}\cos \left( \omega \ \tau +\frac{\pi }{4}\right) 
}{\sqrt{\alpha _{1}}~\bar{\theta}},  \label{E7a} \\
\sin \eta =\frac{k\cos \left( \omega \ \tau +\frac{\pi }{4}\right) +\alpha
_{1}\sin \left( \omega \ \tau +\frac{\pi }{4}\right) }{\sqrt{\alpha _{1}}~%
\bar{\theta}}
\end{gather}%
and without any kind of approximation we find 
\end{subequations}
\begin{eqnarray}
\cos \left( Q_{20}+\eta \right)  &=&\frac{\sqrt{\alpha _{1}+k}~\cos \left(
\omega \tau \right) }{\bar{\theta}}, \\
\sin \left( Q_{20}+\eta \right)  &=&\frac{\sqrt{\alpha _{1}-k}~\sin \left(
\omega \tau \right) }{\bar{\theta}},
\end{eqnarray}%
and from the initial conditions associated with the elliptical mode (\ref%
{E4b}) it follows that 
\begin{equation*}
\sqrt{\alpha _{1}+k}=\bar{\theta}_{0}\ \ ,\ \ \sqrt{\alpha _{1}-k}=\frac{%
\alpha _{2}}{\bar{\theta}_{0}}
\end{equation*}

Picking up all these results in Eq. (\ref{E2a}) we see that 
\begin{equation}
q_{1}=\bar{\theta}\cos \varphi =\bar{\theta}_{0}\cos \left( \omega \tau
\right) \cos \left( \omega _{a}\tau \right) -\frac{\alpha _{2}}{\bar{\theta}%
_{0}}\sin \left( \omega \tau \right) \sin \left( \omega _{a}\tau \right) ,
\end{equation}%
with a similar result for $q_{2}.$

\begin{enumerate}
\item[2.] In order to deduce Eq. (\ref{2.213})%
\begin{equation}
\left[ 
\begin{array}{c}
q_{1} \\ 
q_{2}%
\end{array}%
\right] =%
\begin{bmatrix}
\cos \left( \Omega _{\text{A}}\tau \right)  & -\sin \left( \Omega _{\text{A}%
}\tau \right)  \\ 
\sin \left( \Omega _{\text{A}}\tau \right)  & \cos \left( \Omega _{\text{A}%
}\tau \right) 
\end{bmatrix}%
\begin{bmatrix}
\bar{\theta}_{0}\cos \left( \omega \tau \right)  \\ 
\left( P_{2}/\bar{\theta}_{0}\right) \sin \left( \omega \tau \right) 
\end{bmatrix}
\label{E8}
\end{equation}%
We take into account that $\Phi =\varphi +\pi /2,$ where $\varphi $ is the
spherical coordinate and $\Phi $ is the Euler angle. The definitions (\ref%
{E2}) are converted into 
\begin{subequations}
\label{E9}
\begin{gather}
q_{1}=\frac{x}{l}=\sin \theta \sin \Phi =\ \bar{\theta}\sin \Phi ,
\label{E9a} \\
q_{2}=\frac{y}{l}=-\sin \theta \cos \Phi =-\bar{\theta}\cos \Phi .
\label{E9b}
\end{gather}%
On the other hand, the approximate solution for the azimuthal angle Eqs. (%
\ref{3.116c}), (\ref{3.119b}) and (\ref{3.222}) are given by 
\end{subequations}
\begin{subequations}
\label{E11}
\begin{gather}
\Phi =\left( \frac{3P_{2}}{8}+\frac{P_{\psi }}{2}\right) \tau
+Q_{20}+\arctan \left\{ \frac{1}{P_{2}}\left[ k+P_{1}\tan \left( Q_{1}+\tau
\right) \right] \right\} ,  \label{E11a} \\
Q_{1}=\frac{1}{8}\left( P_{\psi }^{2}-P_{1}\right) \tau +Q_{10},
\label{E11b} \\
Q_{10}=\frac{\pi }{4}\ ,\ \ Q_{20}=\frac{\pi }{2}-\arctan \left( \frac{%
P_{1}+k}{P_{2}}\right) ,  \label{E11c} \\
P_{1}=\frac{P_{2}^{2}}{2\bar{\theta}_{0}^{2}}+\frac{\bar{\theta}_{0}^{2}}{2}%
\ ,\ \ P_{2}=P_{\Phi }-P_{\psi },  \label{E11d} \\
k=P_{1}^{2}-P_{2}^{2}=\frac{\bar{\theta}_{0}^{2}}{2}-\frac{P_{2}^{2}}{2\bar{%
\theta}_{0}^{2}}.  \label{E11e}
\end{gather}%
which can be expressed as 
\end{subequations}
\begin{equation}
\Phi =\Omega _{\text{A}}\tau +\frac{\pi }{2}-\lambda +\eta ,  \label{E10}
\end{equation}%
where we have used the auxiliary functions%
\begin{gather}
\Omega _{\text{A}}\equiv \frac{3P_{2}}{8}+\frac{P_{\psi }}{2}\ ,\ \ \omega
\equiv 1-\frac{P_{1}}{8}+\frac{P_{\psi }^{2}}{8} \\
\lambda \equiv \arctan \left( \frac{P_{1}+k}{P_{2}}\right) \ ,\ \ \eta
\equiv \arctan \left\{ \frac{1}{P_{2}}\left[ k+P_{1}\tan \left( \omega \tau +%
\frac{\pi }{4}\right) \right] \right\} 
\end{gather}%
from these definitions, it is easy to prove that 
\begin{equation}
\cos \lambda =\frac{P_{2}}{\sqrt{2P_{1}}\sqrt{P_{1}+k}}\ ,\ \ \sin \lambda =%
\frac{\sqrt{P_{1}+k}}{\sqrt{2P_{1}}}
\end{equation}%
\begin{gather}
\cos \eta =\frac{P_{2}\left( \cos \omega \tau -\sin \omega \tau \right) }{%
\bar{\theta}\sqrt{2P_{1}}}, \\
\sin \eta =\frac{\left( P_{1}+k\right) \cos \omega \tau +\left(
P_{1}-k\right) \sin \omega \tau }{\bar{\theta}\sqrt{2P_{1}}}.
\end{gather}%
using these results and Eq. (\ref{E10}) in Eqs. (\ref{E9a}, \ref{E9b}), we
have%
\begin{eqnarray*}
q_{1} &=&\bar{\theta}\sin \Phi =\sqrt{P_{1}+k}\cos \left( \Omega _{\text{A}%
}\tau \right) \cos \omega \tau -\sqrt{P_{1}-k}\sin \left( \Omega _{\text{A}%
}\tau \right) \sin \omega \tau , \\
q_{2} &=&-\bar{\theta}\cos \Phi =\sqrt{P_{1}+k}\sin \left( \Omega _{\text{A}%
}\tau \right) \cos \omega \tau +\sqrt{P_{1}-k}\cos \left( \Omega _{\text{A}%
}\tau \right) \sin \omega \tau ,
\end{eqnarray*}%
and considering finally the conditions (\ref{E11d}) and (\ref{E11e}) it is
easy to see that%
\begin{equation}
\sqrt{P_{1}+k}=\bar{\theta}_{0}\ ,\ \ \sqrt{P_{1}-k}=\frac{P_{2}}{\bar{\theta%
}_{0}},
\end{equation}%
which proves Eq. (\ref{E8}).

\item[3.] Now, to prove the canonical character of the linear transformation
in (\ref{3.118}), we introduce a more general linear transformation $(Q,\
P)\rightarrow \left( \Gamma ,\ \Pi \right) $ whose definition is%
\begin{equation}
Q_{i}=f_{i}\left( \Pi \right) \tau +\Gamma _{i}\ ,\ \ P_{i}=\Pi _{i},
\label{linear canonical general}
\end{equation}%
where $f_{i}$ is any polynomic function of the momenta $\Pi _{k}$. The
canonical character of this family of transformations follows from the
Poisson brackets of the variables $Q_{i},P_{i}\ $with respect to the new
variables $\Gamma _{k},\Pi _{k}$%
\begin{equation}
\left[ Q_{i},\ P_{j}\right] _{\Gamma _{k},~\Pi _{k}}=\frac{\partial Q_{i}}{%
\partial \Gamma _{k}}\frac{\partial P_{j}}{\partial \Pi _{k}}-\frac{\partial
Q_{i}}{\partial \Pi _{k}}\frac{\partial P_{j}}{\partial \Gamma _{k}}=\delta
_{ij}
\end{equation}
\end{enumerate}


\begin{thebibliography}{99}
\bibitem{Lagrange} Lagrange, J. L.: \emph{Méchanicque Analitique}. Veuve
Desaint, Paris (1788)

\bibitem{Poisson} Poisson, S. D.: \emph{Sur un cas particulier du mouvement
de rotation des corps pesans}. Journal de I'École Polytechnique. 16, 247-267
(1813)

\bibitem{Golubev} Golubev, V. V.: $\emph{Lectures\ on\ Integration\ of\
Equations\ of\ Motion\ of\ a\ Rigid\ Body\ about\ a\ Fixed\ Point}$. Israeli
Program for Scientific Translations, Israel (1960)

\bibitem{Leimanis} Leimanis, E.: \emph{The General Problem of the Motion of
Coupled Rigid Bodies about a Fixed Point}. Springer Tracts in Natural
Philosophy Vol. 7. Springer Verlag, Berlin (1965)

\bibitem{Johansen} K. F. Johansen and T. R. Kane. \emph{A simple description
of the motion of a spherical Pendulum}. J. Appl. Mech., 36, 76-82. (1969)

\bibitem{Miles} J. W. Miles. \emph{Stability of forced oscillations of a
spherical pendulum}. Q. Appl. Mat., 20, 21-32. (1962)

\bibitem{Hemp} G. W. Hemp., and P. R. Sethna. \emph{The effect of
high-frecuency support oscillation on the motion of a spherical pendulum}.
J. Appl. Mech., 31, 351-354. (1964)

\bibitem{Airy} G. B. Airy, \emph{On the vibration of a free pendulum in an
oval differing little from a straight line}. R. Astron. Soc. XX, 121--130
(1851).(http://home.t01.itscom.net/allais/whiteprior/airy/
airyprecession.pdf)

\bibitem{Olsson} M. G. Olsson, \emph{The precessing spherical pendulum}, Am.
J. Phys. 46 (1978), 1118.

\bibitem{Synge} J. Synge and B. Griffith, \emph{Principles of Mechanics},
McGraw-Hill, New York, 1959

\bibitem{Gusev} A. V. Gusev and M. P. Vinogradov, \emph{Angular velocity of
rotation of the swing plane of a spherical pendulum with anisotropic
suspension}, \textit{Meassurement Techniques}, Vol \textbf{36}, N°10, (1993).

\bibitem{Goldstein} H. Goldstein, C. Poole and J. Safko, \emph{Classical
Mechanics, 3rd Ed. Addison-Wesley (2002)}.

\bibitem{Allais} Allais, M.: \emph{The Allais effect and my experiment with
the paraconical pendulum 1954-1960}. Memories prepared for the NASA, Paris
(1999), 167 pp

\bibitem{Goodey} T. J. Goodey et al., \emph{Correlated anomalous effects
observed during the August 1st 2008 solar eclipse}, Journal of Advanced
Research in Physics 1(2), 021007 (2010).

\bibitem{Olenici} D. Olenici, V. A. Popescu, and B. Olenici, \emph{A
confirmation of the Allais and Javerdan-Rusu-Antonescu effects during the
solar eclipse from 22 September 2006, and the quantization behavior of
pendulum}, Proceedings of the 7th European meeting of the Society for
Scientific Exploration, (2007).

\bibitem{Brizard} A. J. Brizard. \emph{A primer on elliptic functions with
applications in classical mechanics}. arXiv:0711.4064;\ (2007).

\bibitem{Krylov} Bogolyubov, N. and Mitropolsky, Y.: \emph{Asymptotic
methods in theory on nonlinear oscillations}, Ed. Gordon Breachs, New York
(1961).

\bibitem{Whittaker} E. T. Whittaker, \emph{A Treatise on the Analytical
Dynamics of Particles and Rigid Body, 4 ed. Dover, New York (1944).}

\bibitem{Bogolyubov} N. Bogolyubov and Y. Mitropolsky, \emph{Asymptotic
methods in theory on nonlinear oscillations, Ed. Gordon Breachs, New York
(1961).}
\end{thebibliography}
\end{document}